\documentclass[%
reprint,
superscriptaddress,
amsmath,amssymb,
aps,
pra,
]{revtex4-2}

\usepackage{graphicx}
\usepackage{dcolumn}
\usepackage{bm}
\usepackage{amsmath, amssymb}
\usepackage{makecell}
\usepackage{multirow}
\usepackage{float}
\usepackage{color}

\makeatletter

\newcommand{\Rmnum}[1]{\expandafter\@slowromancap\romannumeral #1@}
\makeatother   

\begin{document}                                                                
	\preprint{APS/123-QED}
	
	\title{Asymptotically circularly polarized bound states in the continuum} 

\author{Nan Zhang}
\email{nan25@vt.edu}
\affiliation{Department of Mathematics, City University of Hong Kong, Kowloon, Hong Kong, China
}
\affiliation{Bradley Department of Electrical and Computer Engineering, Virginia Tech, Blacksburg, Virginia 24061, USA
}

\author{Ya Yan Lu}%
\affiliation{Department of Mathematics, City University of Hong Kong, Kowloon, Hong Kong, China
}

\date{\today}

\begin{abstract}
	We study a class of bound states in the continuum (BICs) in all-dielectric periodic structures, near which resonant states approach ideal circularly polarized states (CPSs). We term these BICs {\em asymptotically circularly polarized BICs} ({\em acp}-BICs) and identify two types: single-angle and all-angle. Single-angle {\em acp}-BICs permit convergence to left- or right-handed CPSs only along a single momentum-space direction, whereas all-angle {\em acp}-BICs exhibit convergence to CPSs of a single handedness throughout the entire momentum space, rendering them exceptionally promising for chiral optical applications. 
	We reveal that the existence of {\em acp}-BICs is underpinned by total reflection of circularly polarized waves. 
	Moreover, all-angle {\em acp}-BICs qualify as super-BICs, with uniform nearby polarization being an intrinsic property. 
	In addition,  a bifurcation theory is developed to analyze the emergence of genuine CPSs from {\em acp}-BICs
	under $C_{2}$-symmetric structural perturbations. 
	Our results suggest {\em acp}-BICs as a platform for singular and chiral optical responses in all-dielectric systems.
	
\end{abstract}

%
%

\maketitle
Photonic bound states in the continuum (BICs) 
feature infinite quality ($Q$) factor 
and act as polarization vortex centers in momentum space~\cite{Neumann1929PZ,Friedrich85PRA,Lee12PRL,Hsu13Nature,Zhen14PRL,JZi18PRL,Alu18NP,Hsu16NRM,Kivshar2023PU,Kang2023}, 
enabling applications in lasing~\cite{Kodigala17Nature,Yuri21NC,Rong23NM,Cui25NP}, quantum optics~\cite{santiago2022resonant,rivera2023creating}, and vortex beam generation~\cite{JZi20NP,QSong20Science}, etc. 
More recently, their potential for chiral optics has also attracted considerable attention~\cite{Yuri20PRL,Alu21PRL,Liu21PRL,Qiu22NC,Hamdi23PRB,Yuri23Nature,Pura24,JZi24PRL}, 
including applications in enhanced circular dichroism~\cite{Yuri20PRL,Yuri23Nature,Zagaglia23OL,Kang25PRL}, chiral lasing and emission~\cite{Dixon21PRL,Yuri22Science,lim2023maximally}. 
A central objective is to exploit BICs to generate high-$Q$ circularly polarized states (CPSs), 
which serve as key building blocks for chiral optical responses~\cite{Liu21PRL,Yuri22Science,Yuri23Nature}.
One practical route is to break certain structural symmetries, 
thereby converting a BIC into a pair of CPSs with opposite handedness~\cite{JZi19PRL,Yoda20PRL,CWQiu2021PRL}. 
However, such CPSs typically occur at off-$\Gamma$ points, 
limiting their use in intrinsically chiral applications~\cite{Kang25PRL}. 
It is possible to tune a CPS toward the $\Gamma$ point to realize a chiral quasi-BIC, 
but this generally comes at the cost of a reduced $Q$ factor~\cite{Yuri23Nature,Kang25PRL}.

An alternative approach is to seek special BICs that host nearly circularly polarized states in their vicinity~\cite{JZi24PRL,Kang25PRL}. 
With structural symmetries preserved, these strongly chiral states coexist with the BIC 
and thereby retain high-$Q$ factor. 
Such BICs can be realized in a $C_{6}$-symmetric magneto-optic structure, 
arising from the splitting of doubly degenerate BICs as an external magnetic field 
breaks time-reversal (${\cal T}$) symmetry~\cite{JZi24PRL}.
In this case, nearly CPSs are present throughout the momentum-space neighborhood of the BIC.
For structures with only $C_2$ symmetry, CPSs can be tuned to merge with a BIC, 
leading to pronounced chirality in selected regions of momentum space~\cite{Kang25PRL}.

Here we introduce the concept of {\em asymptotically circularly polarized} ({\em acp}) BICs in dielectric periodic structures. 
An {\em acp}-BIC is defined by a limiting process in which nearby resonant states converge to ideal CPSs along some directions in momentum space. 
In $C_2$-symmetric structures~\cite{Sakoda2005Book}, we identify two types: single-angle and all-angle {\em acp}-BICs, where the resonant states approach left-handed (L) or right-handed (R) CPSs along a single direction or along all directions, respectively. 
All-angle {\em acp}-BICs are particularly significant, as the presence of high-$Q$ nearly CPSs pervading the momentum space endows them with exceptional potential for chiral photonic applications, such as chiral lasing and emission.
The BICs reported in~\cite{JZi24PRL,Kang25PRL} can also be regarded as {\em acp}-BICs.
In particular, the BIC in~\cite{JZi24PRL} corresponds to an all-angle {\em acp}-BIC. 
Distinctively, our work identifies all-angle {\em acp}-BICs in ordinary dielectric structures, 
without the need to break ${\cal T}$ symmetry~\cite{JZi24PRL}.

We next establish the physical criterion for {\em acp}-BICs, 
namely total reflection of circularly polarized incident waves at the BIC frequency and in-plane wavevector. 
Because BICs are decoupled from external excitation, 
the core scattering studies of light-matter interaction have long focused on finite-$Q$ quasi-BICs. 
Nevertheless, the scattering problem associated with BICs is indispensable for identifying {\em acp}-BICs.
We further reveal that all-angle {\em acp}-BICs are also super-BICs, 
for which such uniform polarization is an intrinsic property. 
Super-BICs have been extensively studied for supporting ultrahigh-$Q$ resonances, 
yet the nearby polarization property uncovered here has not been addressed previously~\cite{Yuan17PRA,Zhen19Nature,Kang21PRL,Yuri21NC,Bulgakov23PRB,Le2024PRL,Zhang25PRL}.
Finally, we demonstrate that under structural perturbations preserving $C_{2}$ symmetry, 
only {\em acp}-BICs can bifurcate into exact CPSs. 
This bifurcation can be regarded as the reverse process of a CPS-BIC merging. 
The resulting theory provides a foundation for understanding the symmetry-preserving emergence of CPSs, 
in contrast to the symmetry-breaking scenarios reported previously~\cite{JZi19PRL,Yoda20PRL}.

We first provide a rigorous definition of {\em acp}-BICs.
We consider a dielectric structure embedded in air, periodic in the $xy$ plane and exhibiting mirror symmetry with respect to $z$. 
Our focus is on a resonant state near a nondegenerate BIC.
Assume that only the zeroth diffraction order in air is propagating. 
Exploiting the up-down symmetry, half of the structure can be replaced by an electric or magnetic mirror, 
depending on the field parity in $z$. 
The resonant state radiates into air with far-field polarization vector ${\bm d}$,
which is projected onto the $sp$ plane for analysis,
where the basis vectors ${\hat s}$ and ${\hat p}$ are defined with respect to the BIC.
Let $\mathbb{S}_m$ ($m=0,1,2,3$) denote the Stokes parameters induced from $d_s={\bm d}\cdot\hat{s}$ and $d_p={\bm d}\cdot\hat{p}$.
The degree of circular polarization is defined as $\chi=\mathbb{S}_3/\mathbb{S}_0$.
For in-plane wavevectors ${\bm \kappa}$ and ${\bm \kappa}_*$ of the resonant state and the BIC, 
we write ${\bm\kappa}={\bm\kappa}_*+\delta/L\,{\bm n}$, 
where $\delta>0$ is a small parameter, $L$ a characteristic length scale, and ${\bm n}=(\cos\theta,\sin\theta)$ with $\theta\in [-\pi,\pi)$. 
We call a BIC an {\em acp}-BIC if $\chi(\delta,\theta)\to \pm 1$ as $\delta\to 0$ for some $\theta$.

To characterize resonant states near a BIC, 
a perturbation theory has been developed to capture 
the asymptotic behavior of $Q$ factor and polarizations~\cite{Zhang25PRL,NanSM}. 
We summarize the key results and employ them to study {\em acp}-BICs.
A Bloch mode with in-plane wavevector ${\bm \kappa}$ satisfies 
the eigenequation ${\cal L}({\bm \kappa}){\bm u} = \omega^2 \varepsilon({\bf r}) {\bm u}$, 
where ${\cal L}$ is an operator derived from Maxwell’s equations, 
${\bm u}({\bf r})$ is a periodic electric-field amplitude, 
and $\omega$ is the angular frequency. 
Near a BIC $(\omega_*, {\bm u}_*, {\cal L}_*)$, 
we expand a resonant state $(\omega, {\bm u}, {\cal L})$ in power series of $\delta = |{\bm \kappa} - {\bm \kappa}_*|L$.
As $\delta \to 0$, the polarization vector satisfies ${\bm d}(\delta, \theta)/\delta \to {\bm d}_1(\theta)$, 
where ${\bm d}_1 = d_{1s} \hat{s} + d_{1p} \hat{p}$ is the first-order correction.
It was shown that ${\bm d}_1(\theta)$ depends solely on the BIC and scattering states ${\bm v}_*^s$ and ${\bm v}_*^p$
orthogonal to ${\bm u}_*$. 
Specifically, ${\bm d}_1 = {\bf S}{\bf U}{\bm n}$, where
\begin{equation}
{\bf S}=\left[
\begin{array}{cc}
	R_{ss}&R_{sp}\\
	R_{ps}&R_{pp}
\end{array}
\right],\;{\bf U}=L^2\left[
\begin{array}{cc}
	U_{sx}&U_{sy}\\
	U_{px}&U_{py}
\end{array}
\right].
\end{equation}
In the above, $R_{\sigma\mu}$ denotes the reflection coefficient and $U_{\mu l} = \langle {\bm v}_*^\mu | {\cal L}_{1l} | {\bm u}_* \rangle$
signifies the coupling between scattering states and the BIC. 
The operator ${\cal L}_{1l}$ is the first-order correction to ${\cal L}$ along ${\bm n}=\hat{l}$ with $l\in\{x,y\}$.

When ${\bm d}_1$ becomes circularly polarized at some angle $\theta$, 
i.e., $d_{1s}=\pm i d_{1p}\neq 0$,
we identify an {\em acp}-BIC. 
It is also possible that an {\em acp}-BIC arises from higher-order corrections when ${\bm d}_1=0$. 
For simplicity, in this paper, we restrict our attention to the former case. 
The existence of such ${\bm d}_1$ imposes constraints on the matrices ${\bf S}$ and ${\bf U}$.
For structures with $C_2{\cal T}$ symmetry we find ${\bf S}{\bf U}=\overline{\bf U}$, 
which implies that ${\bf S}^{1/2}{\bf U}$ is real~\cite{NanSM}. 
The scattering matrix ${\bf S}$, being unitary and symmetric~\cite{NanSM}, 
admits the spectral decomposition ${\bf S}={\bf Q}^{\sf T}{\bf \Lambda}{\bf Q}$, 
where ${\bf \Lambda}=\mathrm{diag}(\lambda_1,\lambda_2)$ with $|\lambda_i|=1$ 
and ${\bf Q}$ is a real orthogonal matrix.
Define
\begin{equation}\label{thetavecw}
	{\bm w} := {\bf Q} {\bf S}^{1/2} {\bf U} {\bm n},
\end{equation}
so that ${\bm d}_1^{\sf T} {\bm d}_1 = {\bm w}^{\sf T} {\bf \Lambda} {\bm w}$.
Hence,
${\bm d}_1$ is circularly polarized $\Leftrightarrow$ $\lambda_1=-\lambda_2$ 
and there exists a direction ${\bm n}$ such that ${\bm w}\propto(1,\,1)$ or $(1,\,-1)$. 
The first condition indicates that ${\bf S}$ is traceless,
corresponding to perfect reflection under circularly polarized illumination, 
with $R_{ss} = -R_{pp}$ and $R_{sp} = R_{ps}$. 
This behavior is distinct from that of ordinary mirrors, 
which reverse the handedness of circular polarization upon reflection.

Under the second condition,
if ${\bf U}$ is nonsingular, 
Eq.~\eqref{thetavecw} can be solved directly for ${\bm n}$ with ${\bm w} \propto (1,\,1)$ and $(1,\,-1)$. 
In particular, 
if ${\bm w} \propto (1,\,1)$ corresponds to LCP ${\bm d}_1$, 
then ${\bm w} \propto (1,\,-1)$ corresponds to RCP ${\bm d}_1$, 
and vice versa. 
Owing to the $C_{2}$ symmetry, it suffices to consider $\theta \in [-\pi/2,\pi/2)$ 
and to identify its dual angle with the same $\theta$.
Thus, there exist two distinct angles $\theta_l,\,\theta_r$ such that ${\bm d}_1$ are LCP and RCP, respectively. 
Consequently, we realize a single-angle {\em acp}-BIC, characterized by traceless ${\bf S}$ and nonsingular ${\bf U}$, 
as its nearby resonances approach ideal L- and R-CPSs along $\theta_l$ and $\theta_r$, respectively.

While such single-angle {\em acp}-BICs already host nearly CPSs, 
their presence is typically confined to a narrow angular range. 
Conversely, all-angle {\em acp}-BICs feature single-handed nearly CPSs extending across momentum space, 
making them particularly powerful for chiral applications. 
Writing ${\bf U}$ in column form $[{\bm \xi}_1,\,{\bm \xi}_2]$, 
we obtain ${\bm d}_1=\cos\theta\,\overline{\bm \xi}_1+\sin\theta\,\overline{\bm \xi}_2$ since ${\bf S}{\bf U}=\overline{\bf U}$. 
The condition for ${\bm d}_1$ to be circularly polarized with identical handedness at all $\theta$ 
requires $\overline{\bm \xi}_1$ and $\overline{\bm \xi}_2$ to be linearly dependent, 
with one of them circularly polarized of the same handedness. 
In this case ${\bf U}$ becomes singular, 
and there exists a unique angle $\theta_s$ at which ${\bm d}_1=0$, while ${\bm d}_1 \neq 0$ elsewhere~\cite{Zhang25PRL,NanSM}.
The asymptotic behavior of $\chi$ along $\theta_s$ relies on the high-order corrections.
The resonant states near an all-angle {\em acp}-BIC then converge to an L- or R-CPS, 
depending on the handedness of $\overline{\bm \xi}_1$, 
along all directions except $\theta_s$.

We also have shown that ${\bm d}_1 \neq 0$ corresponds to $Q \sim 1/\delta^2$, 
while ${\bm d}_1 = 0$ corresponds to ultrahigh-$Q$ resonances with $Q \sim 1/\delta^4$ at least~\cite{Zhang25PRL,NanSM}. 
All-angle {\em acp}-BICs are therefore super-BICs with anisotropic asymptotic scaling of $Q$ factor. 
Moreover, any super-BIC with nonzero singular ${\bf U}$--regardless of whether ${\bf S}$ is traceless--exhibits polarization uniformity in its vicinity: 
the resonant states converge to a common polarization defined by $\overline{\bm \xi}_1$ or $\overline{\bm \xi}_2$.

\begin{figure}[htbp]
	\centering
	\includegraphics[scale=0.45]{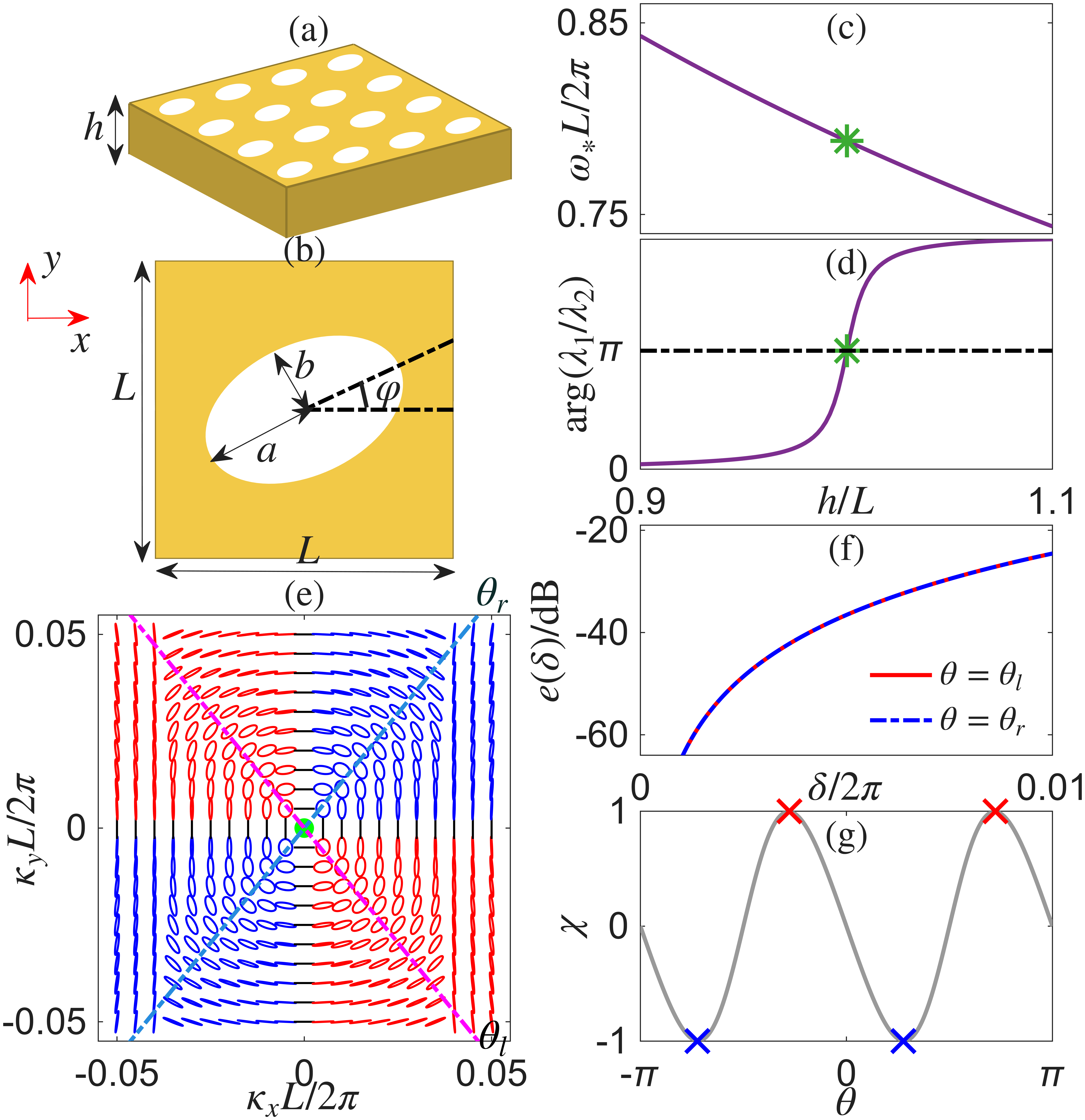}
	\caption{
		(a) Schematic of the PhC slab with a square lattice of elliptic air holes. 
		(b) Top view of a unit cell. 
		(c) and (d) Frequency and phase difference $\arg(\lambda_1/\lambda_2)$ of at-$\Gamma$ BICs as a function of slab height $h$. 
		The green asterisk marks a single-angle {\em acp}-BIC. 
		(e) Polarization pattern of resonant states near the {\em acp}-BIC marked by the green dot. 
		Red and blue ellipses denote elliptic polarization with left- and right-handedness, 
		while red and blue dashed lines indicate the angles $\theta_l$ and $\theta_r$, respectively.
		(f) Error $e(\delta) = ||\chi| - 1|$ quantifying deviation from circular polarization along $\theta=\theta_l,\,\theta_r$. 
		(g) Degree of circular polarization $\chi$  
		as a function of $\theta$ for $\delta/2\pi = 0.01$. 
		Red and blue forks mark the $\theta_l$ and $\theta_r$, respectively.
	}\label{singleangleacpbic}
\end{figure}

To verify our theory, we analyze a photonic crystal (PhC) slab 
with a square lattice of elliptic air holes, as illustrated in Fig.~\ref{singleangleacpbic}(a) and (b).
The slab has a dielectric constant $\varepsilon = 6$, 
corresponding to titanium dioxide at visible frequencies. 
The lattice period, 
slab height, 
semimajor and semiminor axes of the holes, 
and the angle between the semimajor axis and $x$ axis are denoted by $L$, $h$, $a$, $b$, 
and $\varphi$, respectively.
For $a = 0.3L$, $b = 0.15L$, and $\varphi = 0$, 
we vary the slab height $h$, 
compute a family of at-$\Gamma$ symmetry-protected BICs, 
and plot their frequencies as a function of $h$ in Fig.~\ref{singleangleacpbic}(c). 
We also evaluate the scattering matrix ${\bf S}$ associated with these BICs
and plot the phase difference $\arg(\lambda_1/\lambda_2)$ in Fig.~\ref{singleangleacpbic}(d). 
At $h = 1.000L$, we identify a single-angle {\em acp}-BIC characterized by $\lambda_1 = -\lambda_2$ and a nonsingular $\bf U$, 
marked by green asterisks in Fig.~\ref{singleangleacpbic}(c) and (d).
These results also indicate that realizing such a single-angle {\em acp}-BIC is a codimension-1 effect, requiring the tuning of a single parameter.

At the {\em acp}-BIC, we calculate two angles $\theta_l = -0.8698$ and $\theta_r = 0.8698$ 
such that ${\bm d}_1$ is LCP and RCP, respectively.
The relation $\theta_l = -\theta_r$ directly reflects the mirror symmetry of the structure in both $x$ and $y$.
In momentum space, resonant states along $\theta=\theta_l$ and $\theta_r$ approach ideal L-CPS and R-CPS,
respectively. 
The polarization pattern of nearby resonant states is shown in Fig.~\ref{singleangleacpbic}(e),
where nearly CPSs are evident along these directions. 
We further compute the error $e(\delta) = \bigl||\chi|-1\bigr|$ at $\theta=\theta_l,\,\theta_r$, 
and Fig.~\ref{singleangleacpbic}(f) confirms that $e \to 0$ as $\delta \to 0$.
Finally, we plot the degree of circular polarization $\chi$
as a function of $\theta$ at $\delta/2\pi = 0.01$ in Fig.~\ref{singleangleacpbic}(g). 
Near $\theta_l$ and $\theta_r$ highlighted by red and blue forks, 
we observe $|\chi|\approx 1$.

\begin{figure}[htbp]
	\centering
	\includegraphics[scale=0.45]{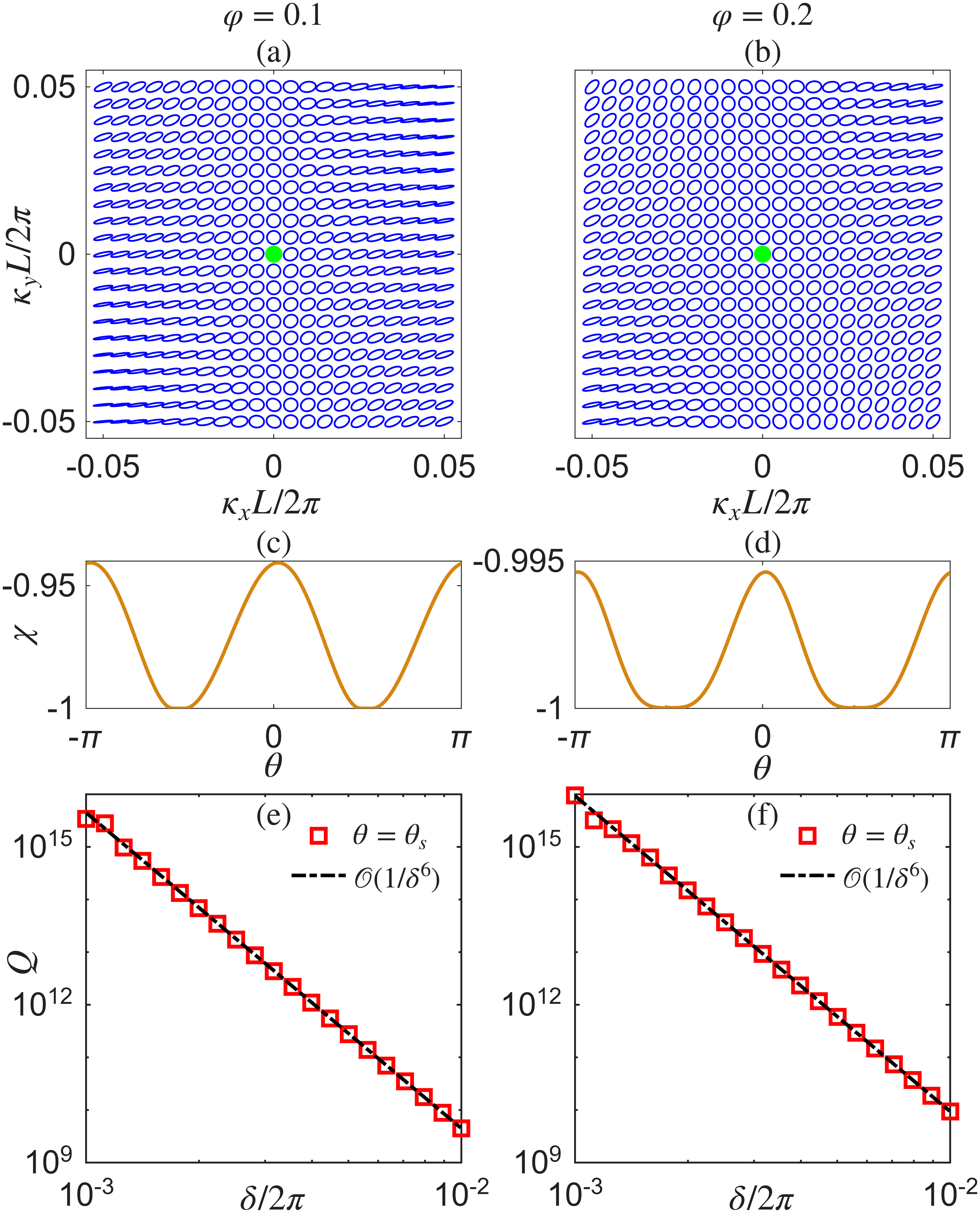}
	\caption{
		All-angle {\em acp}-BICs with nearly CPSs along all directions.  
		(a) and (b) Polarization patterns of resonant states near the all-angle {\em acp}-BICs marked by the green dots at $\varphi = 0.1$ and $0.2$, respectively.  Blue ellipses denote elliptic polarization with right-handedness.
		(c) and (d) Degree of circular polarization $\chi$ as a function of $\theta$ at $\delta/2\pi = 0.01$.  
		(e) and (f) High-order asymptotic scaling of $Q$ factor, $Q = \mathcal{O}(1/\delta^{6})$, at $\theta = \theta_s$.
	}
	\label{allanglebic}
\end{figure}

Note that for the single-angle {\em acp}-BIC,  
nearly CPSs exist only in a narrow region in momentum space,  
as shown in Fig.~\ref{singleangleacpbic}(e) and (g).  
We now search for all-angle {\em acp}-BICs whose nearby resonant states converge to a single-handed CPS along all directions.  
This requires a traceless scattering matrix ${\bf S}$ and a singular ${\bf U}$ containing circularly polarized columns.  
With $C_2{\cal T}$ symmetry, such BICs can be realized by tuning three parameters, indicating a codimension-3 phenomenon.  
We reformulate the problem as a root-finding task and solve it efficiently using Newton’s method~\cite{NanSM}.  
Moreover, for $\varphi=0,\,\pm\pi/4,\,\pm\pi/2$,  
the structure possesses mirror symmetries, and the matrix ${\bf U}$ cannot support circularly polarized columns.  
We therefore exclude these cases.  
At $\varphi = 0.1$ and $0.2$, two {\em acp}-BICs are found at $(a,b,h) = (0.442, 0.265, 0.766)L$ and $(0.408, 0.244, 0.832)L$,  
with $\omega_*L/2\pi=0.859$ and $0.817$, respectively.  
In both cases, $\overline{\bf U}$ possesses RCP columns.  
Here we plot the polarization patterns near the two {\em acp}-BICs in Fig.~\ref{allanglebic}(a) and (b), 
demonstrating that nearly R-CPSs emerge throughout momentum space.  
Figures~\ref{allanglebic}(c) and (d) plot $\chi$ as a function of $\theta$ at $\delta/2\pi = 0.01$,  
where $\chi$ remains close to $-1$ across the entire interval.  
Finally, panels (e) and (f) present the high-order asymptotic behavior of the $Q$ factor along $\theta=\theta_s$,  
verifying that the all-angle {\em acp}-BICs are indeed super-BICs.  

While our examples focus on at-$\Gamma$ BICs,  
the conclusions are general and equally applicable to off-$\Gamma$ cases,  
with representative examples provided in~\cite{NanSM}.  
We next consider structures with higher in-plane rotational symmetry.  
Specifically, we rule out symmetry-protected BICs as {\em acp}-BICs in systems possessing $C_3$, $C_4$, or $C_6$ rotational symmetry~\cite{Sakoda2005Book}.  
By reciprocity and ${\cal T}$ symmetry,  
${\bf S}^{1/2}{\bf U}$ is real and the scattering matrix ${\bf S}$ reduces to a scalar matrix.
Therefore, ${\bf U}$ has linearly polarized columns 
and ${\bm d}_1$ is linearly polarized for all directions $\theta$~\cite{NanSM}.

We have examined the properties of nearly CPSs in the vicinity of an {\em acp}-BIC. 
While resonant states can approach ideal CPSs arbitrarily closely, 
extensive numerical evidence indicates that true CPSs cannot be realized arbitrarily near a BIC in a fixed structure. 
To generate CPSs, one can break the structural $C_{2}$ symmetry, 
which typically causes a BIC to split into a pair of CPSs with opposite handedness~\cite{JZi19PRL,Yoda20PRL}. 
However, under perturbations that preserve $C_{2}$ symmetry, 
a BIC remains robust and generically no CPSs emerge~\cite{Zhen14PRL,Yuan17OL}. 
Here we demonstrate that {\em acp}-BICs constitute an exception, 
as they can bifurcate into exact CPSs.

We present detailed conclusions, with theoretical proofs given in~\cite{NanSM}. 
Focusing on at-$\Gamma$ {\em acp}-BICs, we consider a representative perturbation 
$\Delta\varepsilon({\bf r})=\eta F({\bf r})$, 
where $F({\bf r})$ is confined to the periodic structure and inherits its periodicity. 
Both the original dielectric function $\varepsilon({\bf r})$ and the perturbation profile $F({\bf r})$ preserve $C_2$ symmetry. 
Consequently, a pair of CPSs with identical handedness bifurcates at $\pm\widetilde{\bm{\kappa}}$. 
In the limit $|\eta|\ll 1$, their positions scale as 
$\widetilde{\bm{\kappa}}\sim\sqrt{\eta}\,\widetilde{\bm{\kappa}}_1$, 
exhibiting the characteristic square‑root dependence.
Here $\widetilde{\bm{\kappa}}_1=\widetilde{\kappa}_1{\bm n}$, 
where ${\bm n}$ is a momentum-space direction 
along which the resonant states of the unperturbed structure converge to CPSs. 
The coefficient $\widetilde{\kappa}_1$ depends on both the {\em acp}-BIC and the perturbation profile $F(\mathbf{r})$, 
and is determined by a quadratic relation $\widetilde{\kappa}_1^{2}=\tau_*(F)$~\cite{NanSM}. 
Thus, $\widetilde{\bm{\kappa}}\approx\sqrt{\eta\tau_*}\,{\bm n}$, 
implying that the pair of CPSs exists only for $\eta\tau_*>0$. 
If $\tau_*>0$, a pair of CPSs bifurcates from the {\em acp}-BIC for $\eta>0$, 
whereas for $\eta<0$ the solutions $\sqrt{\eta\tau_*}$ is purely imaginary and no CPSs exist near the BIC. 
An analogous conclusion holds for $\tau_*<0$. 
This constitutes a saddle-node bifurcation of CPSs, 
with the bifurcation point corresponding to the BIC.

For a single-angle {\em acp}-BIC, 
both L- and R-CPS pairs can emerge for either $\eta>0$ or $\eta<0$.
Note that this phenomenon was previously observed in~\cite{Luo23PRA}.
The authors provided a specific numerical example, but our theory is general.
Furthermore, we show that the CPS positions in momentum space satisfy 
$\widetilde{\bm{\kappa}}\approx\sqrt{\eta}\widetilde{\kappa}_{1,l}{\bm n}_{l}$ and 
$\widetilde{\bm{\kappa}}\approx\sqrt{\eta}\widetilde{\kappa}_{1,r}{\bm n}_{r}$, 
where ${\bm n}_{l,r}=(\cos\theta_{l,r},\sin\theta_{l,r})$.

For an all-angle {\em acp}-BIC, in contrast, two CPS pairs of the same handedness can arise~\cite{NanSM}, 
with the handedness determined by the columns of ${\bf U}$. 
Their positions satisfy $\widetilde{\bm{\kappa}}\approx\widetilde{\kappa}_{1}{\bm n}$, 
but unlike the single-angle case, ${\bm n}$ depends on both the BIC and the perturbation profile.
Furthermore, since an {\em acp}-BIC is also a super-BIC, 
off-$\Gamma$ BICs naturally emerge along angle $\theta_{s}$ as a byproduct of the bifurcation process~\cite{Zhang2024OL}.

\begin{figure}[htbp]
	\centering
	\includegraphics[scale=0.45]{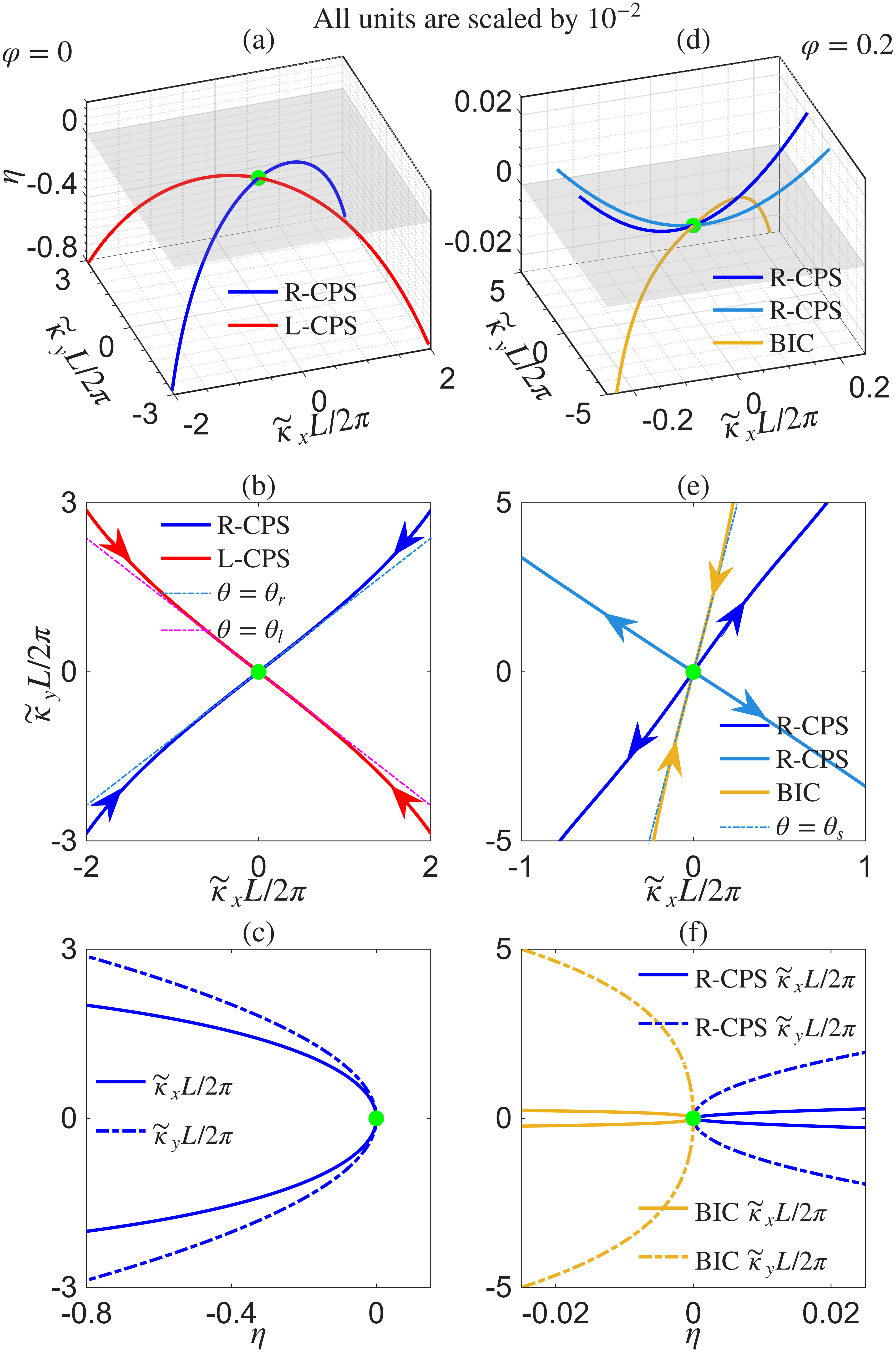}
	\caption{Bifurcation phenomena: CPSs emerging from the single-angle {\em acp}-BIC with $\varphi=0$
		and the all-angle {\em acp}-BIC with $\varphi=0.2$.
		Note that all units are scaled by $10^{-2}$.
		(a) and (d) Trajectories of CPSs in $({\widetilde{\bm \kappa}},\eta)$ space with $\eta=(h-h_*)/L$. 
		The two BICs are marked by green dots. 
		A pair of off-$\Gamma$ BICs also emerge from the all-angle {\em acp}-BIC. 
		(b) and (e) Projections of CPSs and BICs in momentum space, with arrows indicating the direction of increasing $h$. 
		(c) and (f) Bifurcation diagrams for the pair of R-CPSs from the single-angle {\em acp}-BIC (c), 
		and for one pair of R-CPSs together with the off-$\Gamma$ BICs from the all-angle {\em acp}-BIC (f).}
	\label{bifurcation}
\end{figure}

To verify our theory, we compute CPSs emerging from two {\em acp}-BICs discussed above: 
the single-angle {\em acp}-BIC with $h_*=1.000L$ and $\varphi=0$, 
and the all-angle {\em acp}-BIC with $h_*=0.832L$ and $\varphi=0.2$. 
We vary only the height $h$ to locate the CPSs, with the structural perturbation amplitude defined as $\eta=(h-h_*)/L$. 
For the single-angle {\em acp}-BIC, the CPS trajectories in $(\widetilde{\bm \kappa},\eta)$ space and their momentum-space projections are shown in Fig.~\ref{bifurcation}(a) and (b).
As $h$ is decreased from $h_*$ ($\eta<0$), a pair of L-CPSs and a pair of R-CPSs emerge from the BIC, 
with their locations asymptotically aligned along $\theta_l$ and $\theta_r$. 
Conversely, upon increasing $h$ to $h_*$ ($\eta\to 0^-$), these pairs merge back into the BIC along the same directions. 
These directions are also associated with the limiting process: 
in the original structure with $h=h_*$, nearby resonant states converge to ideal CPSs along them. 
Figure~\ref{bifurcation}(c) shows $\widetilde{\kappa}_{x}$ and $\widetilde{\kappa}_{y}$ of the R-CPS pair as functions of $\eta$. 
It is evident that a saddle-node bifurcation occurs, with the BIC representing the bifurcation point. 
As $\eta \to 0^-$, a square-root asymptotic behavior is observed.
Since the momentum-space locations of the L-CPSs are simply the mirror images of those of the R-CPSs, 
they exhibit similar results.

For the all-angle {\em acp}-BIC at $\varphi=0.2$, the trajectory of the CPSs in $(\widetilde{\bm \kappa},\eta)$ space is shown in Fig.~\ref{bifurcation}(d). 
In this case, two pairs of R-CPSs emerge from the BIC as $h$ is increased from $h_*$, 
while no L-CPSs appear. 
In addition, a pair of off-$\Gamma$ BICs emerge since the all-angle {\em acp}-BIC is also a super-BIC,
and their locations are verified to approach asymptotically along $\theta_s$.
The projections in momentum space are shown in Fig.~\ref{bifurcation}(e). 
Figure~\ref{bifurcation}(f) presents $\widetilde{\kappa}_{x}$ and $\widetilde{\kappa}_{y}$ of the off-$\Gamma$ BICs and one pair of R-CPSs as functions of $\eta$. 
Bifurcation phenomena are observed in this case as well. 
For these two {\em acp}‑BICs, we further shows that fixing $h_*$
and varying semiminor axis $b$ to search for CPSs yields the same bifurcation behavior~\cite{NanSM}.

In conclusion, we have reported two types of {\em acp}-BICs in $C_{2}$-symmetric dielectric structures, 
near which resonant states approach ideal L-CPS or R-CPS along either a single direction or all directions in momentum space. 
We showed that the existence of {\em acp}-BICs is rooted in the scattering problem, 
namely the total reflection of a circularly polarized incident wave. 
We further revealed that all-angle {\em acp}-BICs are super-BICs. 
While the scattering problem associated with a BIC and the polarization properties of super-BICs are usually overlooked, 
here we unify them and demonstrate all-angle {\em acp}-BICs, 
which may enable chiral optical applications such as chiral lasing and emission. 
In addition, we established a connection between {\em acp}-BICs and true CPSs, 
and developed a bifurcation theory to quantitatively characterize the emergence of CPSs from these BICs. 
We believe that our results hold significant potential for singular and chiral optics.

The authors acknowledge support from the Research Grants Council of Hong
Kong Special Administrative Region, China (Grant No. CityU 11317622). 

\bibliography{acpbic}

\onecolumngrid

\renewcommand{\thesection}{S\arabic{section}}
\renewcommand{\theequation}{S\arabic{equation}}

\indent

\begin{center}\large
	\textbf{Appendix}
\end{center}

\renewcommand{\thesection}{A\arabic{section}}
\renewcommand{\theequation}{A\arabic{equation}}
\numberwithin{equation}{section}
\renewcommand{\thefigure}{A\arabic{figure}}

	\section{Units, notation, and background}
	This section specifies the units and notation adopted in the paper 
	and provides a concise review of perturbation theory for resonant states near a BIC. 
	We use the following notation: quantities related to a BIC are denoted with a subscript ``$*$''; 
	complex conjugation is indicated by ``$-$''; 
	the transpose of a matrix or vector is denoted by ``${\sf T}$''. 
	We work in Lorentz-Heaviside units with $\varepsilon_0=\mu_0=c=1$. 
	The Maxwell’s equations then read
	\begin{equation}
		\nabla\times {\bm E} = i\omega {\bm H},\quad  \nabla\times {\bm H} = -i\omega \varepsilon({\bf r}){\bm E},
	\end{equation}
	where ${\bf r}=(x, y, z)$, $\omega$ has units of $1/[{\bf r}]$. 
	
	In a dielectric periodic structure embedded in air, 
	a Bloch mode has a quasi-periodic electric field 
	${\bm E}={\bm u}({\bf r})\exp(i{\bm \kappa}\cdot {\bm \rho})$, 
	where ${\bm \rho}=(x,y)$ and ${\bm\kappa}=(\kappa_x,\kappa_y)$. 
	The periodic amplitude ${\bm u}$ satisfies the wave equation 
	\begin{equation}\label{maineq}
		{\cal M}{\bm u}({\bf r})=0,
	\end{equation}
	where ${\cal M}={\cal L}({\bm \kappa})-\omega^2\varepsilon({\bf r})$ is the wave operator and ${\cal L}=(\nabla+i{\bm \kappa})\times(\nabla+i{\bm \kappa})\times$.
	Assume that the structure supports a BIC $(\omega_*,{\bm \kappa}_*,{\bm u}_*)$
	and $\varepsilon({\bf r})$ has up-down mirror symmetry in $z$. 
	For convenience, 
	the lower half of the structure is replaced by an electric or magnetic mirror, depending on the field parity. 
	The BIC is normalized as
	\begin{equation}
		\langle {\bm u}_*|\varepsilon|{\bm u}_*\rangle = \frac{1}{L^3} \int_\Omega \overline{\bm u}_*({\bf r}) \varepsilon({\bf r}) {\bm u}_*({\bf r})\, d{\bf r}=1,
	\end{equation}
	where $\Omega$ denotes a unit cell of the periodic structure.

	Since the wavevector of a resonant state is complex, 
	typically no real polarization plane can be well defined. 
	For a resonant state ($\omega$, ${\bm \kappa}$, ${\bm u}$) near the BIC,
	we therefore project the polarization vector ${\bm d}$ onto the $sp$-plane associated with the BIC. 
	The basis vectors are defined by 
	$\hat{s}=\hat{z}\times\hat{k}_*/|\hat{z}\times\hat{k}_*|$ and $\hat{p}=\hat{k}_*\times {\hat s}$, 
	where $\hat{k}_*$ is the unit vector of ${\bm k}_*=({\bm \kappa}_*,\gamma_*)$ 
	and $\gamma_*=\sqrt{\omega_*^2-|{\bm \kappa}_*|^2}$. 
	If $|\hat{z}\times\hat{k}_*|=0$, we set $\hat{s}=\hat{x}$. 
	The Stokes parameters are
	\begin{equation}\label{stokespara}
		\mathbb{S}_0=|d_{s}|^2+|d_{p}|^2,\;\mathbb{S}_1=|d_{s}|^2-|d_{p}|^2,\;\mathbb{S}_2=2\mbox{Re}(d_s\overline{d}_p),\;\mathbb{S}_3=-2\mbox{Im}(d_s\overline{d}_p),
	\end{equation}
	where $d_s={\bm d}\cdot\hat{s}$ and $d_p={\bm d}\cdot\hat{p}$.  
	For R-CPSs and L-CPSs, we have $d_s=id_p$ and $d_s=-id_p$, 
	with $\mathbb{S}_1=\mathbb{S}_2=0$ and $\chi=\mathbb{S}_3/\mathbb{S}_0=-1$ and $1$, respectively.
	
	At the BIC frequency and in-plane wavevector, 
	we define two output and input channels of the system by
	\begin{equation}
		{\bf s}_{\rm out} = \eta\hat{s}e^{i\gamma_* z},\quad {\bf p}_{\rm out} = \eta\hat{p}e^{i\gamma_* z},\quad {\bf s}_{\rm in} = {\cal C}_2{\cal T}{\bf s}_{\rm out},\quad {\bf p}_{\rm in} = {\cal C}_2{\cal T}{\bf p}_{\rm out},
	\end{equation}
	where $\eta=0.5i/\gamma_*L$ is a normalization constant,
	${\cal T}$ is time-reversal operator given by $ {\cal T}{\bm u}=\overline{\bm u}$,
	and ${\cal C}_2$ is a $180^\circ$ rotation operator around the $z$ axis given by
	\begin{equation}
		{\cal C}_2{\bm u}=\left[
		\begin{array}{r}
			-u_x(-x,-y,z)\\
			-u_y(-x,-y,z)\\
			u_z(-x,-y,z)            
		\end{array}
		\right].
	\end{equation}
	Since the structure is lossless, we can obtain ${\bf S}{\bf S}^\dagger={\bf S}^\dagger{\bf S}={\bf I}_2$ directly,
	where ${\bf I}_2$ is $2\times 2$ identity matrix.
	Moreover, in $C_2$-symmetric structures,
	we have ${\bf S}={\bf S}^{\sf T}$.
	In fact, considering a scattering state ${\bm v}$ with input amplitudes ${\bf a}=(a_s,a_p)$ and output amplitudes ${\bf b}=(b_s,b_p)$,
	we have ${\bf S}{\bf a}={\bf b}$ and ${\bf S}\overline{\bf b}=\overline{\bf a}$ since ${\cal C}_2{\cal T}{\bm v}$ is also a scattering state.
	Therefore, ${\bf S}\overline{\bf S}=\overline{\bf S}{\bf S}={\bf I}_2$.
	The normalized coefficients of the R‐CPS and L‐CPS waves in the output channels (${\bf s}_{\rm out}$, ${\bf p}_{\rm out}$) are 
	$(1,-i)/\sqrt{2}$ and $(1,i)/\sqrt{2}$, respectively,
	while in the input channels they are $(1,i)/\sqrt{2}$ and $(1,-i)/\sqrt{2}$, respectively. 
	If ${\bf S}$ is traceless,
	perfect reflection of a circularly polarized incident wave occurs.
	
	We briefly review the perturbation theory developed in~\cite{Zhang25PRL}.
	We parameterize the wavevector as ${\bm \kappa} = {\bm \kappa}_*+\delta/L\,{\bm n}$
	with $0<\delta\ll 1$.
	The first-order corrections to $(\omega,{\bm u},{\cal L})$ are $\omega_1$, ${\bm u}_1$ and ${\cal L}_1$, 
	respectively, where 
	\begin{equation}
		{\cal L}_{1}=\frac{i}{L}\left[\left(\nabla+i{{\bm \kappa}_*}\right)\times{\bm n}+{\bm n}\times\left(\nabla+i{{\bm \kappa}_*}\right)\right]\times.
	\end{equation}
	The operator ${\cal L}_1$ can also be decomposed as ${\cal L}_{1}={\cal L}_{1x}\cos\theta+{\cal L}_{1y}\sin\theta$.
	The zeroth-order perturbation equation ${\cal M}_*{\bm u}_*=0$ is the wave equation related to the BIC, 
	and the first-order perturbation equation is
	\begin{equation}\label{LEq1}
		{\cal M}_*{\bm u}_1={\bm f}_1:=2\omega_*\omega_1\varepsilon ({\bf r}) {\bm u}_*-{\cal L}_1{\bm u}_*.
	\end{equation}
	Since ${\bm f}_1$ vanishes at infinity and can be regarded as a finite-energy source, 
	the far-field pattern of ${\bm u}_1$ is a propagating plane wave with the polarization vector ${\bm d}_1=d_{1s}\hat{s}+d_{1p}\hat{p}$. 
	We introduce two scattering states ${\bm v}_*^s$ and ${\bm v}_*^p$,
	corresponding to ${\bf a}=(1,0)$ and $(0,1)$,
	to build a linear relationship between ${\bm d}_1$ and ${\bm n}$.
	We also shift the scattering states such that they are orthogonal with the BIC, i.e.,
	$\langle{\bm v}_*^\mu|\varepsilon|{\bm u}_*\rangle=0$. 
	Using the Lorentz reciprocity theorem, the polarization vector ${\bm d}_1$ can be obtained from
	$\langle{\bm v}_*^\mu|{\cal M}_*|{\bm u}_1\rangle=\langle{\bm v}_*^\mu|{\bm f}_1\rangle$,
	yielding 
	\begin{equation}
		{\bm d}_1={\bf S}{\bf U}{\bm n}.
	\end{equation}
	To streamline notation, we make no distinction between ${\bm d}_1$ and $(d_{1s}, d_{1p})$.
	
	\section{Proof of ${\bf S}{\bf U}=\overline{\bf U}$}
	In this section, we show that ${\bf S}{\bf U}=\overline{\bf U}$ 
	for BICs in structures with $C_{2}{\cal T}$ symmetry, 
	as well as for symmetry-protected BICs in structures with $C_{n}{\cal T}$ symmetry 
	($n=3,4,6$). 
	We begin with the former case.
	The BIC is scaled such that ${\cal C}_2{\cal T}{\bm u}_*={\bm u}_*$. 
	Since $[{\cal C}_2{\cal T},\,{\cal L}_{1}]=0$, 
	it follows that
	$\langle {\cal C}_2{\cal T}{\bm v}|{\cal L}_{1}|{\cal C}_2{\cal T}{\bm u}\rangle=\overline{\langle{\bm v}|{\cal L}_{1}|{\bm u}\rangle}$.
	Moreover,
	\begin{equation}
		{\cal C}_2{\cal T}\left[
		\begin{array}{c}
			{\bm v}_*^s\\
			{\bm v}_*^p
		\end{array}
		\right]=\overline{\bf S}\left[
		\begin{array}{c}
			{\bm v}_*^s\\
			{\bm v}_*^p
		\end{array}
		\right].
	\end{equation}
	Therefore,
	\begin{equation}
		\overline{\bf U} = L^2\left[
		\begin{array}{c|c}
			\langle {\cal C}_2{\cal T}{\bm v}_*^s|{\cal L}_{1x}|{\cal C}_2{\cal T}{\bm u}_*\rangle&\langle {\cal C}_2{\cal T}{\bm v}_*^s|{\cal L}_{1y}|{\cal C}_2{\cal T}{\bm u}_*\rangle\\
			\hline
			\langle {\cal C}_2{\cal T}{\bm v}_*^p|{\cal L}_{1x}|{\cal C}_2{\cal T}{\bm u}_*\rangle&\langle {\cal C}_2{\cal T}{\bm v}_*^p|{\cal L}_{1y}|{\cal C}_2{\cal T}{\bm u}_*\rangle
		\end{array}
		\right]={\bf S}{\bf U}.
	\end{equation}
	
	Next, we turn to the latter case. 
	Here the BIC is scaled with ${\cal T}{\bm u}_*={\bm u}_*$,
	and
	\begin{equation}
		{\cal T}\left[\begin{array}{c}
			{\bm v}_*^s\\
			{\bm v}_*^p
		\end{array}\right]=-\overline{\bf S}\left[\begin{array}{c}
			{\bm v}_*^s\\
			{\bm v}_*^p
		\end{array}\right].
	\end{equation}
	We also have $\langle {\cal T}{\bm u}|{\cal L}_{1l}|{\cal T}{\bm v}\rangle=-\overline{\langle {\bm u}|{\cal L}_{1l}|{\bm v}\rangle}$.
	Thus,
	\begin{equation}
		-\overline{\bf U} = L^2\left[
		\begin{array}{c|c}
			\langle {\cal T}{\bm v}_*^s|{\cal L}_{1x}|{\cal T}{\bm u}_*\rangle&\langle {\cal T}{\bm v}_*^s|{\cal L}_{1y}|{\cal T}{\bm u}_*\rangle\\
			\hline
			\langle {\cal T}{\bm v}_*^p|{\cal L}_{1x}|{\cal T}{\bm u}_*\rangle&\langle {\cal T}{\bm v}_*^p|{\cal L}_{1y}|{\cal T}{\bm u}_*\rangle
		\end{array}
		\right]=-{\bf S}{\bf U}
	\end{equation}
	Hence, in both cases we obtain ${\bf S}{\bf U}=\overline{\bf U}$.
	
	\section{Nearly linear polarizations surrounding at-$\Gamma$ BICs}
	In this section, we show that resonant states exhibit nearly linear polarizations
	in the vicinity of at-$\Gamma$ BICs with nonsingular ${\bf U}$ in structures with $C_n$ symmetry, $n=3,4,6$.
	Equivalently, we demonstrate that ${\bm d}_1$ is linearly polarized for all $\theta$.
	Such BICs arise from the {\em A}‑irreducible representation in $C_n$ symmetry, $n=3,4,6$, and 
	from the {\em B}‑irreducible representation in $C_4$ symmetry.
	
	We show that linearly polarized ${\bm d}_1$
	follows directly from the scalar form of
	${\bf S}=\exp{(i\phi_S)}{\bf I}_2$.
	Since ${\bf S}{\bf U}=\overline{\bf U}$, we obtain
	${\bf S}^{1/2}{\bf U}=\exp{(i\phi_S/2)}{\bf U}$, which is real.
	Thus every element of ${\bf U}$ carries a phase of either $-\phi_S/2$ or $-\phi_S/2+\pi$.
	Writing ${\bf U}=[{\bm \xi}_1,\,{\bm \xi}_2]$,
	both ${\bm \xi}_1$ and ${\bm \xi}_2$ are linearly polarized,
	and hence ${\bm d}_1=\cos\theta\,\overline{\bm \xi}_1+\sin\theta\,\overline{\bm \xi}_2$ is linearly polarized.
	
	We show that ${\bf S}$ is a scalar matrix. Let ${\bf R}(\theta)$ denote the rotation matrix with $\theta=2\pi/n$ ($n=3,4,6$).  
	The space spanned by ${\bm v}_*^s$ and ${\bm v}_*^p$ is invariant under ${\cal R}\in\{{\cal C}_3,{\cal C}_4,{\cal C}_6\}$.
	We then have ${\bf R}{\bf S}={\bf S}{\bf R}$.
	Writing
	\begin{equation}
		{\bf S}=\begin{bmatrix}a&b\\ c&d\end{bmatrix},
	\end{equation}
	we find $a=d$, $c=-b$,
	i.e.,
	${\bf S}=a{\bf I}_2+b{\bf J}_2$ with
	\begin{equation}
		{\bf J}_2=\left[\begin{array}{rr}
			0&-1\\
			1& 0
		\end{array}\right].
	\end{equation}                    
	Lorentz reciprocity requires ${\bf S}={\bf S}^{\mathsf T}$. 
	Since ${\bf J}_2$ is skew-symmetric, 
	$b=0$, and thus ${\bf S}=a{\bf I}_2$ with $a=\exp(i\phi_S)$, as claimed.
	
	\noindent {\textbf{\textit {Remark}}}: For structures invariant under $C_{2v}$, 
	the scattering matrix takes the diagonal form
	\begin{equation}
		{\bf S} = \begin{bmatrix}a&\\&b\end{bmatrix}.
	\end{equation}
	
	We show that at-$\Gamma$ BICs with nonsingular ${\bf U}$ arise from the {\em A}‑irreducible representation in $C_n$ symmetry, $n=3,4,6$, and 
	from the {\em B}‑irreducible representation in $C_4$ symmetry.
	A direct calculation shows
	\begin{equation}
		{\cal R}\,{\cal L}_1({\bm n})\,{\cal R}^{-1}={\cal L}({\bf R}{\bm n}).
	\end{equation}
	For ${\bm n}=\hat{x}$ and $\hat{y}$, 
	we have a matrix form
	\begin{equation}
		{\cal R}\!\begin{pmatrix}{\cal L}_{1x}\\[2pt]{\cal L}_{1y}\end{pmatrix}\!{\cal R}^{-1}
		= {\bf R}^{\sf T}\begin{pmatrix}{\cal L}_{1x}\\[2pt]{\cal L}_{1y}\end{pmatrix}.
	\end{equation}
	Combined with $
	\left<{\bm v}_{*}^\mu|{\cal L}_{1l}|{\bm u}_*\right>=\left<{\cal R}{\bm v}_{*}^\mu{\cal R}|{\cal L}_{1l}|{\bm u}_*\right>,
	$
	we have
	\begin{equation}
		{\bf U}=L^2\left[
		\begin{array}{c|c}
			\langle {\cal R}{\bm v}_*^s|{\cal R}{\cal L}_{1x}|{\bm u}_*\rangle&\langle {\cal R}{\bm v}_*^s|{\cal R}{\cal L}_{1y}|{\bm u}_*\rangle\\
			\hline
			\langle {\cal R}{\bm v}_*^p|{\cal R}{\cal L}_{1x}|{\bm u}_*\rangle&\langle {\cal R}{\bm v}_*^p|{\cal R}{\cal L}_{1y}|{\bm u}_*\rangle
		\end{array}
		\right]=\tau_{\cal R}{\bf R}^{\sf T}{\bf U}{\bf R},
	\end{equation}
	where ${\cal R}{\bm u}_*=\tau_{\cal R}{\bm u}_*$.
	Representing ${\bf U}$ with Pauli matrices $\sigma_i$, 
	we find:
	
	1. For $\tau_{\cal R}=1$ in $C_n$ symmetry ($n=3,4,6$, {\em A}‑irreducible representation), ${\bf U}=a\sigma_1^2+b(i\sigma_2)$ with $\det{\bf U}=a^2+b^2$.
	
	2. For $\tau_{\cal R}=-1$ in $C_4$ symmetry ({\em B}‑irreducible representation), ${\bf U}=a\sigma_3+b\sigma_1$ with $\det{\bf U}=-a^2-b^2$.
	
	\noindent In both cases, $\exp(i\phi_S/2){\bf U}$ is real, implying that $a\overline{b}$ is real.
	In generic case, $a$ and $b$ are not both zero and then $\det{\bf U}$ is nonzero.
	
	\noindent {\textbf{\textit {Remark}}}: In $C_n$ symmetry with $n=2,6$, 
	at-$\Gamma$ BICs with $\tau_{\cal R}=-1$ ({\em B}-irreducible representation)
	yield ${\bf U}=0$ automatically.
	These are super-BICs since ${\bm d}_1=0$ and $Q\sim 1/\delta^4$ at least for all $\theta$.
	In $C_{2v}$ symmetry,
	at-$\Gamma$ BICs with $\tau_{\cal R}=1$ ({\em A}-irreducible representation) yield
	\begin{equation}
		{\bf U} = \begin{bmatrix}a&\\&b\end{bmatrix}\text{ or }{\bf U} = \begin{bmatrix}&a\\b&\end{bmatrix}.
	\end{equation}
	
	\section{Uniform polarizations near a super-BIC}
\begin{figure}[htbp]
		\centering
		\includegraphics[scale=0.7]{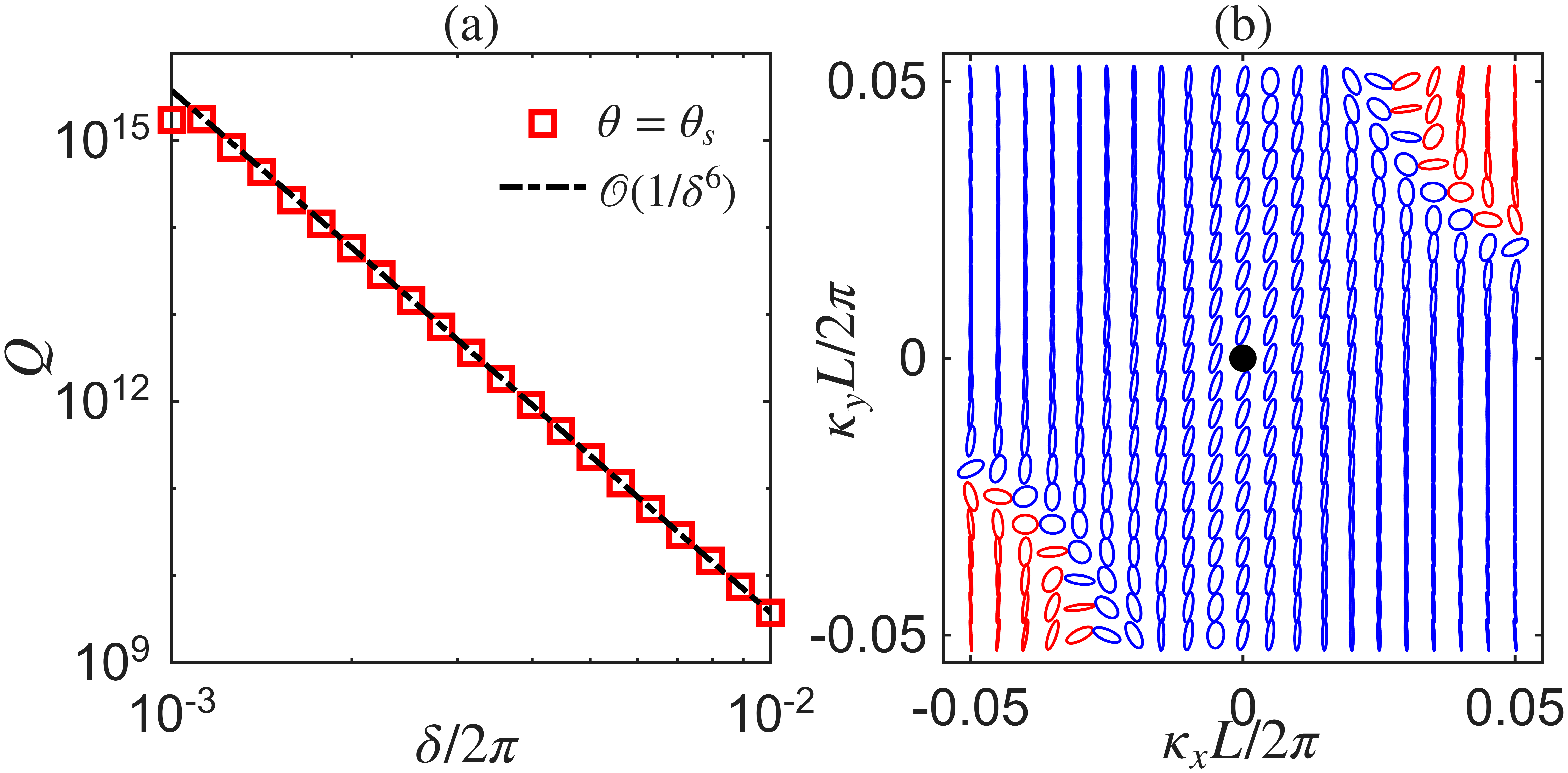}
		\caption{
			(a) Asymptotic behavior of $Q$ factor along $\theta_s$.
			(b) Polarization pattern  of resonant states near the super-BIC.
		}
		\label{superbicmomentum}
	\end{figure}

	In this section, we give an example on the uniform polarization 
	near a super-BIC with a singular nonzero ${\bf U}$.
	With
	$(a,b,h)=(0.4,0.27,0.767)L$ and $\varphi=0.2$,
	we find such a super-BIC at $\omega_* L/2\pi=0.850$
	with ${\bm d}_1=0$ for $\theta_s=1.498$. 
	The asymptotic behavior of $Q$ factor along $\theta=\theta_s$
	and polarization pattern in momentum space are shown in Fig.~\ref{superbicmomentum}, 
	confirming that resonant states near the super-BIC indeed possess uniform polarization.
	
	
	\section{Numerical methods}
	
	We reformulate the search for {\em acp}-BICs as a root-finding problem
	${\bm G}({\bm p})=0$, where ${\bm p}$ denotes the structural parameters,
	and solve it using Newton’s method.
	For single-angle {\em acp}-BICs,
	\begin{equation}
		{\bm G}({\bm p})=
		\begin{bmatrix}
			\lambda_{\rm min}(\widehat{\bf S})\\
			\mathrm{Tr}({\bf S})
		\end{bmatrix},
	\end{equation}
	where $\widehat{\bf S}$ is the total scattering matrix including evanescent channels,
	$\lambda_{\rm min}$ is its allest-magnitude eigenvalue,
	and $\mathrm{Tr}({\bf S})$ is the trace of ${\bf S}$.
	For all-angle {\em acp}-BICs,
	\begin{equation}
		{\bm G}({\bm p})=
		\begin{bmatrix}
			\lambda_{\rm min}(\widehat{\bf S})\\
			U_{sx}\pm iU_{px}\\
			U_{sy}\pm iU_{py}
		\end{bmatrix},
	\end{equation}
	since ${\bf S}{\bf U}=\overline{\bf U}$ and ${\bm d}_1=\overline{\bf U}{\bm n}$.
	The matrices $\widehat{\bf S}$ and ${\bf U}$ are constructed using the Fourier Modal Method~\cite{liu2012s4}.
	
	\section{All-angle {\em acp}-BICs in parameter space}
	
	Searching for an all-angle {\em acp}-BIC requires tuning three structural parameters.
	This follows because one column of ${\bf U}$ must be circularly polarized
	and simultaneously $\det{\bf U}=0$.
	Without loss of generality, we assume the first column $(U_{sx},U_{px})$ is right-handed circularly polarized,
	which requires tuning two parameters such that
	\begin{equation}
		\mathrm{Re}(U_{sx}-iU_{px})=0, \qquad \mathrm{Im}(U_{sx}-iU_{px})=0 .
	\end{equation}
	The condition $\det{\bf U}=0$ requires tuning one additional parameter,
	because ${\bf S}^{1/2}{\bf U}$ is real and therefore
	\begin{equation}
		\det({\bf S}^{1/2}{\bf U})=0
	\end{equation}
	imposes a single constraint.
	
\begin{figure}[htbp]
		\centering
		\includegraphics[scale=0.6]{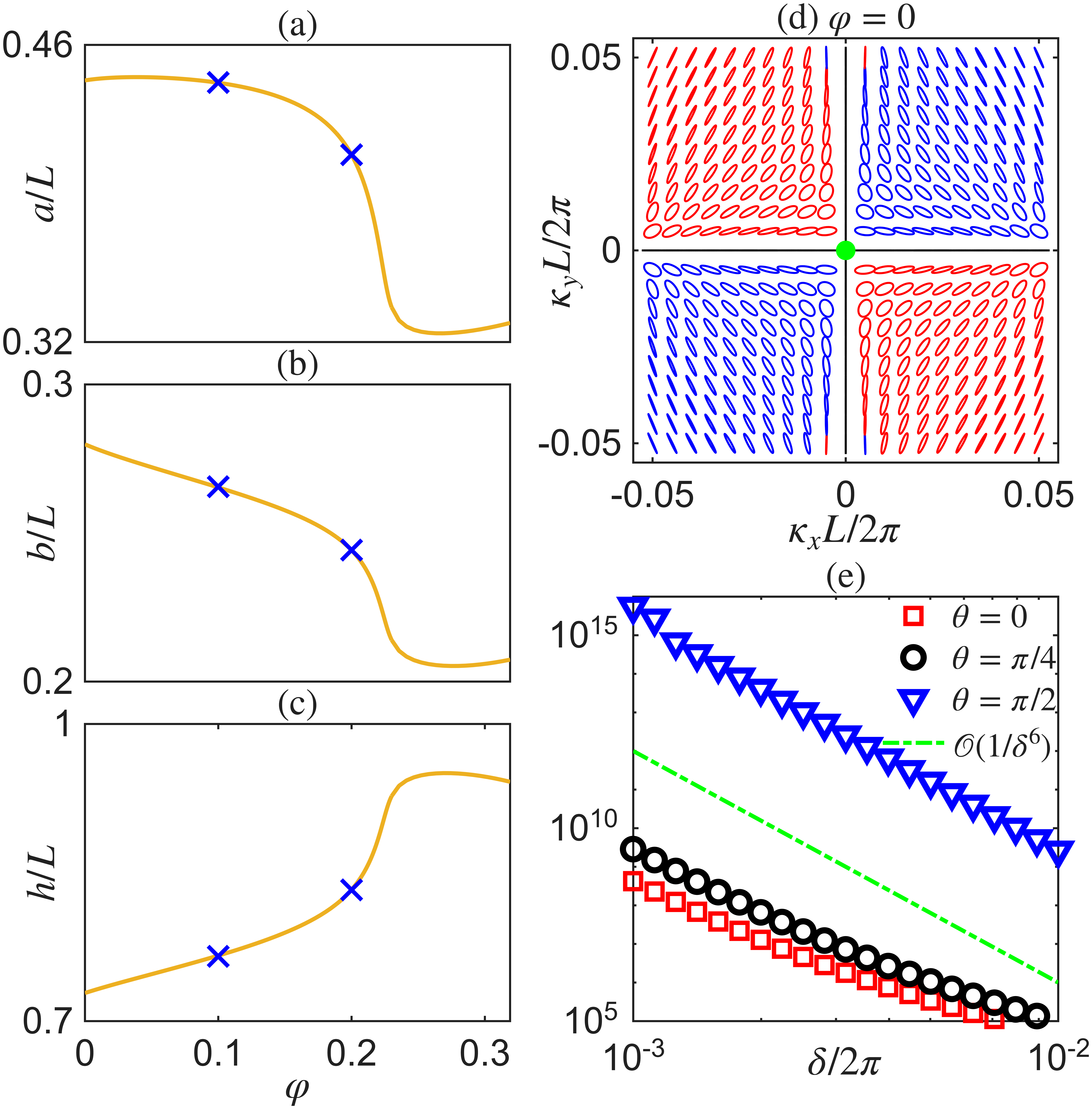}
		\caption{(a)--(c) Values of $a$, $b$, and $h$ versus $\varphi$ for all-angle {\em acp}-BICs.
			Blue forks mark the two all-angle {\em acp}-BICs discussed in the paper.
			(d) Polarization pattern of resonant states near the BIC at $\varphi=0$.
			(e) Higher-order asymptotic behavior of $Q$ factor along $\theta=0,\pi/4,\pi/2$.}
		\label{acpbicparameter}
	\end{figure}

	We study all-angle {\em acp}-BICs in parameter space. 
	For the structure considered in the paper, 
	we obtain a family by varying $\varphi$ and searching for ${\bm p}=(a,b,h)$ such that ${\bm G}({\bm p})=0$. 
	The results are shown in Fig.~\ref{acpbicparameter}(a)--(c).
	Two all-angle {\em acp}-BICs discussed in the paper are marked by blue forks.
	We also find a limiting point at $\varphi=0$. 
	In this case, the structure possesses $C_{2v}$ symmetry, 
	so ${\bf U}$ takes either diagonal or anti-diagonal form, as given in Sec.~A3. 
	However, since ${\bf U}$ must contain circularly polarized columns on the parameter manifold supporting all-angle {\em acp}-BICs, 
	${\bf U}$ vanishes in this case. 
	The asymptotic behavior of polarizations depends on higher-order corrections. 
	We plot the polarization pattern of resonant states near the BIC in Fig.~\ref{acpbicparameter}(d). 
	Owing to reflection symmetry, it is evident that along $k_x=0$ or $k_y=0$ the polarization is linear. 
	Thus this BIC represents only a limiting case of all-angle {\em acp}-BICs, and not a true all-angle {\em acp}-BIC. 
	Nevertheless, this observation provides a natural starting point for the search for all-angle {\em acp}-BICs, 
	namely BICs satisfying $\mathrm{Tr}({\bf S})=0$ and ${\bf U}=0$.
	In addition, since ${\bf U}=0$ at this point, the BIC is a super-BIC, 
	with higher-order asymptotic scaling of $Q$ factor appearing in all directions [see Fig.~\ref{acpbicparameter}(e)].
	
	\section{Bifurcation theory for CPSs}
	
	In the paper we presented detailed conclusions on the bifurcation of CPSs from {\em acp}-BICs. 
	In this section, we provide the theoretical proof. 
	Assume that a CPS $(\widetilde{\omega},\widetilde{\bm \kappa},\widetilde{\bm u})$ emerges from such a BIC in a perturbed structure with dielectric function
	\begin{equation}\label{perdie}
		\widetilde{\varepsilon}({\bf r})=\varepsilon({\bf r})+\eta F({\bf r}).
	\end{equation}
	For sufficiently all $\eta$, we expand the CPS in a power series of $\zeta$:
	\begin{align}
		\label{peromega}
		\widetilde{\omega}&=\omega_*+\zeta\widetilde{\omega}_1+\zeta^2\widetilde{\omega}_2+\cdots,\\
		\label{perbeta}
		\widetilde{\bm \kappa}&={\bm \kappa}_*+\zeta\widetilde{\bm \kappa}_1+\zeta^2\widetilde{\bm \kappa}_2+\cdots,\\
		\label{peru}
		\widetilde{\bm u}&={\bm u}_*+\zeta\widetilde{\bm u}_1+\zeta^2\widetilde{\bm u}_2+\cdots.
	\end{align}
	Here $\zeta={\eta}^{1/P}$, with the integer $P$ to be determined. 
	We also expand the operator $\widetilde{\cal L}$ associated with the CPS as
	\begin{equation}\label{perL}
		\widetilde{\cal L}={\cal L}_*+\zeta\widetilde{\cal L}_1+\zeta^2\widetilde{\cal L}_2+\cdots.
	\end{equation}
	Let $\widetilde{\bm \kappa}_j=\widetilde{\kappa}_j{\bm n}_j$ and ${\kappa}_j=\widetilde{\kappa}_jL$. 
	Then the operator $\widetilde{\cal L}_j$ can be written as
	\begin{equation}
		\widetilde{\cal L}_j={\kappa}_j{\cal L}_1({\bm n}_j)+\sum_{m=0}^{j-1}{\kappa}_m{\kappa}_{j-m}{\cal L}_2({\bm n}_m,{\bm n}_{j-m}),
	\end{equation}
	where the operator function ${\cal L}_1({\bm n})$ is defined in Sec.~A1, and
	\begin{equation}
		{\cal L}_2({\bm n}_1,{\bm n}_2) = -\frac{1}{L^2}{\bm n}_1\times {\bm n}_2\times.
	\end{equation}
	Here and in the following, we use the subscript $0$ to denote quantities of the BIC.
	Substituting Eqs.~\eqref{perdie} and \eqref{peromega}--\eqref{perL} into the wave equation $\widetilde{\cal M}\widetilde{\bm u}=0$ satisfied by the CPS, and collecting terms of order ${\cal O}(\zeta^j)$, we obtain
	\begin{equation}\label{perEqs}
		{\cal M}_*\widetilde{\bm u}_j=\widetilde{\bm f}_j({\bf r}):=-\sum_{m=1}^{j}\widetilde{\cal L}_m\widetilde{\bm u}_{j-m}+\sum_{m=1}^{j}W_m\widetilde{\bm u}_{j-m},
	\end{equation}
	where
	\begin{equation}
		W_m=\sum_{l=0}^{m}\widetilde{\omega}_l\widetilde{\omega}_{m-l}\varepsilon({\bf r})+\sum_{l=0}^{m-2}\widetilde{\omega}_l\widetilde{\omega}_{m-2-l}F({\bf r}).
	\end{equation}
	The far-field pattern of the CPS takes the form
	\begin{equation}\label{ffcps}
		\widetilde{\bm u}\sim \widetilde{\bm d} e^{i\widetilde{\gamma} z},\quad z\to \infty,
	\end{equation}
	where $\widetilde{\gamma}=\sqrt{\widetilde{\omega}^2-|\widetilde{\bm \kappa}|^2}$. 
	Expanding Eq.~\eqref{ffcps} in a power series yields
	\begin{equation}
		\widetilde{\bm d} e^{ i\widetilde{\gamma} z}=\left(\zeta \widetilde{\bm d}_1+\zeta^2\widetilde{\bm d}_2+\cdots\right)\left[1+\zeta g_1 (z)+\zeta^2g_2(z)+\cdots\right]e^{i\gamma_* z},
	\end{equation}
	where $g_j (z)$ can be expressed explicitly using
	$\widetilde{\bm \kappa}_1,\widetilde{\omega}_1,\cdots,\widetilde{\bm \kappa}_{j},\widetilde{\omega}_{j}$. 
	Accordingly, the far-field pattern of $\widetilde{\bm u}_j$ can be extracted as
	\begin{equation}\label{ffpatternuj}
		\widetilde{\bm u}_j\sim{\bm \psi}_j(z) e^{i\gamma_* z},\quad z\rightarrow\infty,\quad {\bm \psi}_j(z) = \widetilde{\bm d}_{j}+\sum_{m=1}^{j-1}\widetilde{\bm d}_{m} g_{j-m}(z).
	\end{equation}
	Thus, we show that there exists an integer $P$ such that, for every $j \geq 1$, 
	a real $\zeta^j\widetilde{\bm \kappa}_j$ can be found for which Eq.~\eqref{perEqs} admits a solution $\widetilde{\bm u}_j$ with circularly polarized ${\bm \psi}_j$.

	We solve Eq.~\eqref{perEqs} directly to verify the conditions. 
	Our analysis focuses on at-$\Gamma$ single-angle {\em acp}-BICs, while the theory can be generalized to other cases. 
	In the following, we assume the CPS to be an R-CPS.
	The case of an L-CPS can be analyzed analogously. 
	The $C_2$ symmetry enforces $P=2$, as a pair of CPSs bifurcates at $\pm\widetilde{\bm \kappa}$. 
	For simplicity, we solve only the leading-order case, while higher-order terms can be obtained analogously. 
	
	For $j=1$, we have ${\bm \psi}_1=\widetilde{\bm d}_1$ and
	\begin{equation}
		\widetilde{\bm f}_1 = -{\kappa}_1{\cal L}_{1}{\bm u}_*+2\omega_*\widetilde{\omega}_1\varepsilon({\bf r}){\bm u}_*.
	\end{equation}
	The solvability condition $\langle {\bm u}_*|\widetilde{\bm f}_1\rangle=0$, together with the anti-commutativity ${\cal C}_2{\cal L}_{1}=-{\cal L}_{1}{\cal C}_2$, gives $\widetilde{\omega}_1=0$. 
	Therefore, $\widetilde{\bm f}_1={\kappa}_1{\bm f}_1$, $\widetilde{\bm u}_1={\kappa}_1{\bm u}_1$, and $\widetilde{\bm d}_1={\kappa}_1{\bm d}_1$, where ${\bm f}_1$, ${\bm u}_1$, and ${\bm d}_1$ are defined in Sec.~A1. 
	Using the scattering states ${\bm v}_*^\mu$ ($\mu\in\{s,p\}$), we obtain $\widetilde{\bm d}_1={\kappa}_1{\bf S}{\bf U}{\bm n}$. 
	For ${\bm n}={\bm n}_r$, $\widetilde{\bm d}_1$ corresponds to RCP.
	
	To determine ${\kappa}_1$, we consider higher-order terms. 
	For $j=2$, the right-hand side is
	\begin{equation}
		\widetilde{\bm f}_2=-\kappa_{2}{\cal L}_{1}({\bm n}_2){\bm u}_*+{\kappa}_1^2{\bm f}_{21}+2\omega_*\widetilde{\omega}_2\varepsilon({\bf r}){\bm u}_*
		+\omega_*^2F({\bf r}){\bm u}_*,
	\end{equation}
	with
	\begin{equation}
		{\bm f}_{21}=-{\cal L}_2({\bm n}_r,{\bm n}_r){\bm u}_*-{\cal L}_1({\bm n}_r){\bm u}_1.
	\end{equation}
	The solvability condition $\langle{\bm u}_*|\widetilde{\bm f}_2\rangle=0$ gives $2\omega_*\widetilde{\omega}_2=\kappa_1^2W_{21}+W_{22}$, where $W_{21}=-\langle{\bm u}_*|{\bm f}_{21}\rangle$ and $W_{22}=-\omega_*^2\langle{\bm u}_*|F|{\bm u}_*\rangle$. 
	The far-field polarization vector is ${\bm \psi}_2=\widetilde{\bm d}_2+\widetilde{\bm d}_1 g_1(z)$. 
	Since $\widetilde{\omega}_1=0$, $g_1=0$, and thus ${\bm \psi}_2=\widetilde{\bm d}_2$. 
	We require that the projection of $\widetilde{\bm d}_2$ onto the $sp$-plane be circularly polarized, i.e., $\widetilde{d}_{2s}=i\widetilde{d}_{2p}$, where $\widetilde{d}_{2s}=\widetilde{\bm d}_2\cdot\hat{s}$ and $\widetilde{d}_{2p}=\widetilde{\bm d}_2\cdot\hat{p}$.
	Via the Lorentz reciprocity theorem and symmetry conditions, 
	the relation $\langle {\bm v}_*^\mu|{\cal M}_*|\widetilde{\bm u}_2\rangle=\langle {\bm v}_*^\mu|\widetilde{\bm f}_2\rangle$
	also gives rise to a linear constraint between $\widetilde{\bm \kappa}_2$ and $\widetilde{\bm d}_2$:
	\begin{equation}
		\left[\begin{array}{c}
			\widetilde{d}_{2s}\\
			\widetilde{d}_{2p}
		\end{array}\right]={\kappa}_2{\bf S}{\bf U}{\bm n}_2.
	\end{equation}
	We then have $\widetilde{\bm\kappa}_2=\widetilde{\kappa}_2{\bm n}_r$, 
	where $\widetilde{\kappa}_2$ is a constant to be determined. 
	We can rewrite $\widetilde{\bm f}_2$ as
	\begin{equation}
		\widetilde{\bm f}_2 = {\kappa}_2{\bm f}_1+{\kappa}_1^2{\bm f}_2+W_{22}\varepsilon({\bf r}){\bm u}_*+\omega_*^2F{\bm u}_*,
	\end{equation}
	where ${\bm f}_2={\bm f}_{21}+W_{21}\varepsilon({\bf r}){\bm u}_*$. 
	The solution $\widetilde{\bm u}_2$ can be formally written as $\widetilde{\bm u}_2=\kappa_2{\bm u}_1+\kappa_1^2{\bm u}_2+{\bm u}_{21}$, where ${\cal M}_*{\bm u}_2={\bm f}_2$ and ${\cal M}_*{\bm u}_{21}=W_{22}\varepsilon({\bf r}){\bm u}_*+\omega_*^2F{\bm u}_*$.
	
	We solve for $\kappa_1$ at $j=3$. 
	The right-hand side ${\bm f}_3$ can be written as
	\begin{equation}
		{\bm f}_3=-\kappa_{3}{\cal L}_{1}({\bm n}_3){\bm u}_*+2\omega_*\widetilde{\omega}_3\varepsilon({\bf r}){\bm u}_*
		+\kappa_1^3{\bm f}_{3,1}+\kappa_1{\bm f}_{3,2}(F)+2\kappa_1\kappa_2{\bm f}_{21},
	\end{equation}
	where
	\begin{equation}
		\begin{aligned}
			{\bm f}_{3,1}&=-{\cal L}_{1}{\bm u}_2-{\cal L}_{2}{\bm u}_1+W_{21}\varepsilon({\bf r}){\bm u}_*,\\
			{\bm f}_{3,2}(F)&=-{\cal L}_{1}{\bm u}_{21}+[W_{22}\varepsilon({\bf r})+\omega_*^2F({\bf r})]{\bm u}_1.
		\end{aligned}
	\end{equation}
	According to Eq.~\eqref{ffpatternuj}, the far-field pattern of $\widetilde{\bm u}_3$ is given by ${\bm \psi}_3(z)= \widetilde{\bm d}_{3}+\widetilde{\bm d}_{1} g_2(z)$, where $g_2(z)=iz(2\omega_*\omega_2-\widetilde{\kappa}_1^2)/(2\gamma_*)$. 
	Therefore,
	\begin{equation}
		{\bm \psi}_3(z)=\widetilde{\bm d}_{3}+\kappa_1^3z\frac{i({W}_{21}-1/L^2)}{2\gamma_*}{\bm d}_1+\kappa_1z\frac{iW_{22}}{2\gamma_*}{\bm d}_1.
	\end{equation}
	Notice that ${\bm \psi}_3(z)$ is a polynomial in $z$. 
	Therefore, the improper integral $\langle{\bm v}_{*}^\mu|{\cal M}_*|\widetilde{\bm u}_3\rangle$ typically cannot converge. 
	However, considering the integral in a finite domain, we obtain
	\begin{equation}
		L^3\langle\phi_h{\bm v}_{*}^\mu|{\cal M}_*|\widetilde{\bm u}_3\rangle = \int_{S} G_{3\mu}(x,y,h)\, dS=L^3\langle\phi_h{\bm v}_{*}^\mu|{\bm f}_3\rangle,
	\end{equation}
	where $\phi_h\equiv 1$ in $S\times(0,h)$ and $\phi_h\equiv 0$ elsewhere, and
	\begin{equation}
		G_{3\mu}({\bf r})=\hat{z} \cdot \left[\widetilde{\bm u}_3 \times \nabla\times \overline{\bm v}_{*}^\mu -  \overline{\bm v}_{*}^\mu \times \nabla\times \widetilde{\bm u}_3\right].
	\end{equation}
	In the limit $h\to \infty$, the divergent contributions on both sides must cancel. 
	Therefore, we obtain a linear equation
	\begin{equation}
		\left[\begin{array}{c}
			\widetilde{d}_{3s}\\
			\widetilde{d}_{3p}
		\end{array}\right]
		={\bf S}\left[{\bf U}(\kappa_3{\bm n}_3)+\kappa_1\left(\kappa_1^2{\bf c}_{3,1}+{\bf c}_{3,2}\right)\right],
	\end{equation}
	where ${\bf c}_{3,1}$ and ${\bf c}_{3,2}$ can be written down explicitly. 
	Notice that the above formula does not include the unknown $\kappa_2$, 
	since the symmetry conditions ${\cal C}_2{\bm f}_{21}={\bm f}_{21}$ 
	and ${\cal C}_2{\bm v}_*^\mu=-{\bm v}_*^\mu$ hold.
	The condition $\widetilde{d}_{3s}=i\widetilde{d}_{3p}$ induces a real linear system
	\begin{equation}
		{\bf A}(\kappa_3{\bm n}_3)={\kappa}_1\left[{\kappa}_1^2{\bf b}_{3,1}+{\bf b}_{3,2}(F)\right].
	\end{equation}
	The solvability of the above linear system requires
	\begin{equation}
		{\kappa}_1{\bm n}_r^{\sf T}\left({\kappa}_1^2{\bf b}_{3,1}+{\bf b}_{3,2}\right)=0,
	\end{equation}
	since ${\bf A}{\bm n}_r=0$. 
	We then obtain $\widetilde{\kappa}_1=0$ or $\widetilde{\kappa}_1=\sqrt{\tau_*(F)}$, where
	\begin{equation}
		\tau_*=-\frac{1}{L^2}\frac{{\bm n}_r^{\sf T}{\bf b}_{3,2}}{{\bm n}_r^{\sf T}{\bf b}_{3,1}}.
	\end{equation}	
	The case ${\kappa}_1\neq 0$ corresponds to the CPS, 
	while ${\kappa}_1=0$ corresponds to a new at-$\Gamma$ BIC, 
	which connects to the robustness theory of BICs in structures with $C_2$ symmetry.
	Since $\widetilde{\bm \kappa}\approx \sqrt{\eta\tau_*}{\bm n}_r$, 
	the pair of R-CPSs exists only for $\eta\tau_*>0$.
	
	\section{Numerical examples: CPSs emerging from {\em acp}-BICs}
	
\begin{figure}[htbp]
		\centering
		\includegraphics[scale=0.6]{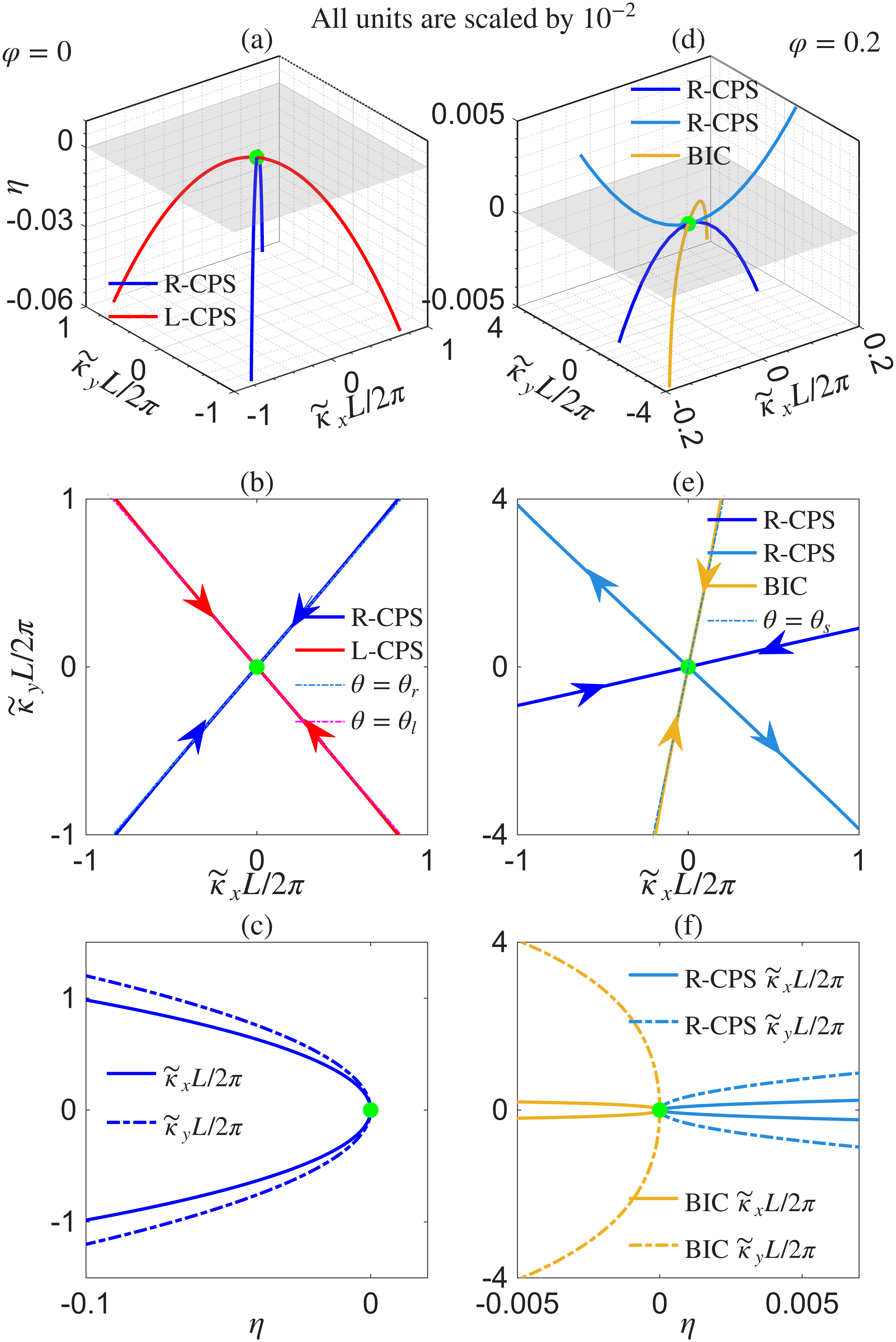}
		\caption{Bifurcation phenomena: CPSs emerging from the single-angle {\em acp}-BIC with $\varphi=0$
			and the all-angle {\em acp}-BIC with $\varphi=0.2$.
			Note that all units are scaled by $10^{-2}$.
			(a) and (d) Trajectories of CPSs in $({\widetilde{\bm \kappa}},\eta)$ space with $\eta=(b-b_*)/L$. 
			The two BICs are marked by green dots. 
			A pair of off-$\Gamma$ BICs also emerge from the all-angle {\em acp}-BIC asymptotically align $\theta_s$. 
			(b) and (e) Projections of CPSs and BICs in momentum space, with arrows indicating the direction of increasing $h$. 
			(c) and (f) Bifurcation diagrams for the pair of R-CPSs from the single-angle {\em acp}-BIC (c), 
			and for one pair of R-CPSs together with the off-$\Gamma$ BICs from the all-angle {\em acp}-BIC (f).}
		\label{smbifurcation}
	\end{figure}

	In this section we vary only the semiminor axis $b$ to compute CPSs and off-$\Gamma$ BICs
	emerging from the two {\em acp}-BICs at $\varphi=0$ and $\varphi=0.2$ discussed in the paper.
	We define $\eta=(b-b_*)/L$ and present the results in Fig.~\ref{smbifurcation}.
	For the single-angle {\em acp}-BIC at $\varphi=0$,
	the CPS locations align asymptotically with $\theta_r$ and $\theta_l$.
	Together with the results in the paper,
	this confirms that the asymptotic directions of CPSs are independent of structural perturbations.
	In contrast, for the all-angle {\em acp}-BIC at $\varphi=0.2$,
	although off-$\Gamma$ BICs also align asymptotically with $\theta_{s}$,
	the asymptotic CPS directions obtained by varying $b$ differ from those obtained by varying $h$,
	indicating their dependence on structural perturbations.
	
	\section{Numerical examples: Off-$\Gamma$ {\em acp}-BICs and a closed loop of CPSs}
	
\begin{figure*}[htbp]
		\centering
		\includegraphics[scale=0.4]{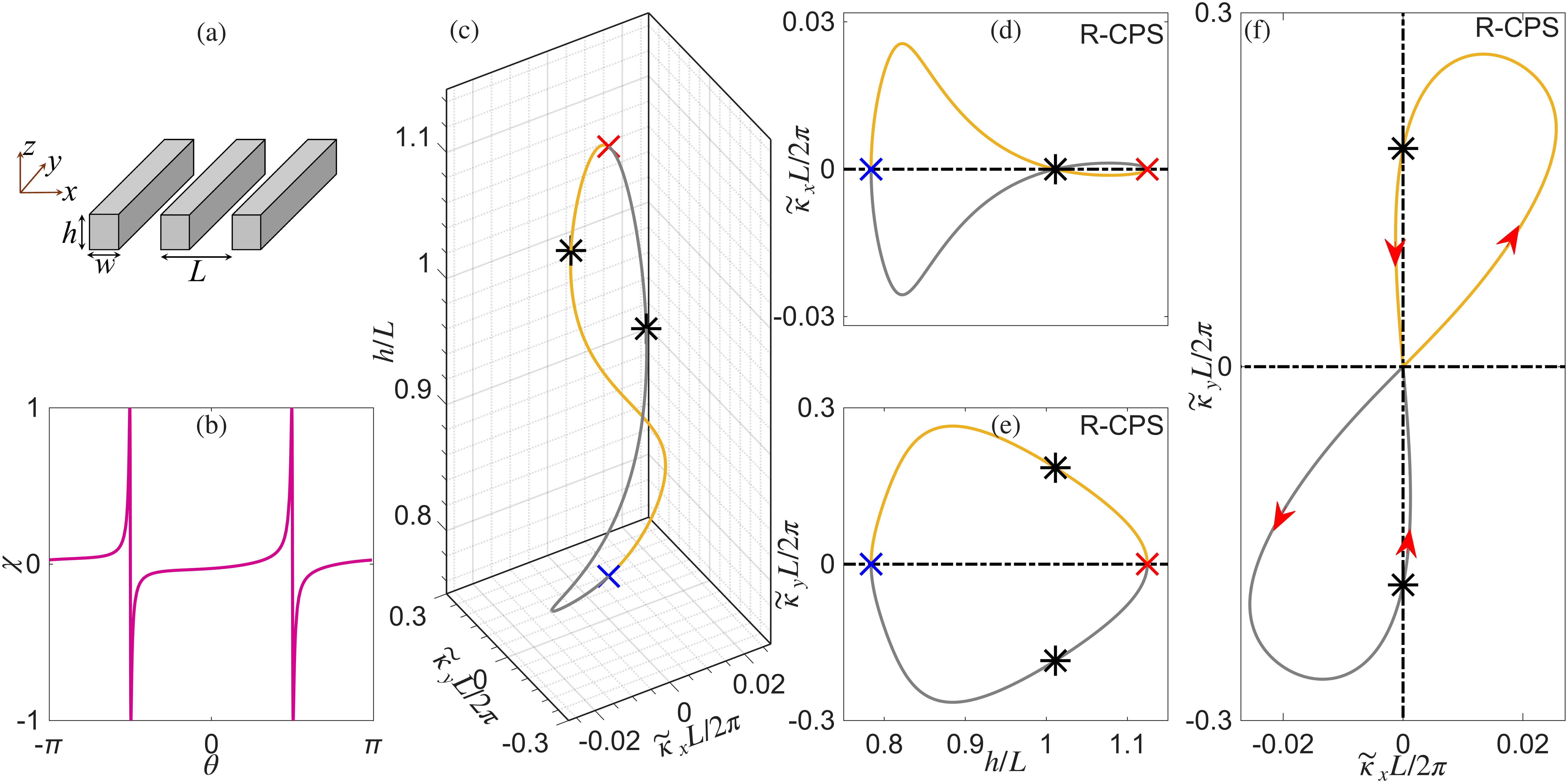}
		\caption{(a) Schematic of the diffraction grating.
			(b) Degree of circular polarization $\chi$ as a function of $\theta$.
			(c) Trajectories of CPSs in $({\widetilde{\bm \kappa}},h)$ space. 
			The two at-$\Gamma$ {\em acp}-BICs are marked by blue and red forks, respectively, 
			and the off-$\Gamma$ {\em acp}-BICs by black asterisks.
			(d)--(f) Projections of the trajectories in $(h,\widetilde{\kappa}_x)$, $(h,\widetilde{\kappa}_y)$,
			and $(\widetilde{\kappa}_x,\widetilde{\kappa}_y)$ planes. 
			Square-root and linearly asymptotic behavior appear near at-$\Gamma$ and off-$\Gamma$ {\em acp}-BICs, respectively.}
		\label{acpbicgrating}
	\end{figure*}

	In this section we compute off-$\Gamma$ {\em acp}-BICs and study the trajectory 
	of CPSs in momentum space as structural parameters are varied. 
	Since the CPSs under consideration may not lie in the vicinity of the BIC, 
	we define the far-field polarization plane of resonant states, 
	referred to as the $sp$-plane induced by the unit vector 
	$\widetilde{\bm k}=\bigl(\widetilde{\bm \kappa}, \sqrt{\mathrm{Re}(\widetilde{\omega})^2 - |\widetilde{\bm \kappa}|^2}\bigr)$. 
	We consider the diffraction grating shown in Fig.~\ref{acpbicgrating}(a), 
	consisting of a periodic array of infinitely long rectangular TiO$_2$ ($\varepsilon=6$) rods in air.
	The period, width, and height of the rods are denoted by $L$, $w$, and $h$, respectively. 
	We fix the width at $w=0.3L$ and obtain an off-$\Gamma$ single-angle {\em acp}-BIC at 
	$h_*=1.011L$, ${\bm \kappa}_*L/2\pi=(0,0.185)$, and $\omega_* L/2\pi=0.735$. 
	As shown in Fig.~\ref{acpbicgrating}(b), 
	$\chi$ plotted versus $\theta$ exhibits $|\chi|\approx 1$ at specific angles. 
	We also obtain two at-$\Gamma$ {\em acp}-BICs at $h_*=0.784L$ and $h_*=1.125L$, 
	with $\omega_* L/2\pi=0.738$ and $0.719$, respectively.
	
	We next compute the CPSs emerging from the three {\em acp}-BICs. 
	Owing to the reflection symmetry of the structure, we consider only R-CPSs. 
	As shown in Fig.~\ref{acpbicgrating}(c), the R-CPSs form a closed loop: 
	as $h$ increases, CPSs emerge from the {\em acp}-BIC marked by a blue fork, 
	cross the off-$\Gamma$ {\em acp}-BIC marked by a black asterisk, 
	and finally merge with the {\em acp}-BIC marked by a red fork. 
	We also plot the projections of this loop in the $(h,\widetilde{\kappa}_x)$, $(h,\widetilde{\kappa}_y)$, 
	and $(\widetilde{\kappa}_x,\widetilde{\kappa}_y)$ planes. 
	A key distinction between the at-$\Gamma$ and off-$\Gamma$ cases lies in the asymptotic behavior of CPS locations. 
	Specifically, CPSs emerging from off-$\Gamma$ {\em acp}-BICs exhibit linear asymptotic behavior.
	Such a phenomenon can be explained by our bifurcation theory with $\zeta=\eta$.

\end{document}